# Increasing Downshifting Luminescence Intensity Through an Extended Active Layer

*Miao Liu, Jinyang Liang\*, Fiorenzo Vetrone\**

Centre Énergie, Matériaux et Télécommunications, Institut National de la Recherche Scientifique, Université du Québec, 1650 boulevard Lionel-Boulet, Varennes, Québec J3X 1P7, CANADA

*Corresponding authors: jinyang.liang@inrs.ca (J.L.), fiorenzo.vetrone@inrs.ca (F.V.)

**Abstract**

The near-infrared (NIR) emission of rare-earth doped nanoparticles (RENPs), known as downshifting luminescence, has been extensively investigated in diverse applications from information technology to biomedicine. In promoting brightness and enriching the functionalities of the downshifting luminescence of RENPs, numerous studies have exploited inert shell to protect rare-earth dopants from surface quenchers. However, internal concentration quenching remains an unsolved puzzle when using higher dopant concentrations of rare-earth ions in an attempt to obtain brighter emission. Following a plethora of research involving core-shell structures, the interface has shown to be controllable, ranging from a well-defined, abrupt boundary to an obscure one with cation intermixing. By utilizing this inter-mixed core-shell property for the first time, we design a new architecture to create a homogeneous double-layer core-shell interface to extend the active layer, allowing more luminescent centers without severe concentration quenching. By systematically deploying the crystallinity of the starting core, shell growth dynamics, and dopant concentrations, the downshifting luminescence intensity of new archictecture achieves a 12-fold enhancement surpassing the traditional core-shell structure. These results provide deeper insight into the potential benefits of the intermixed core-shell structure, offering an effective approach to tackling the internal concentration quenching effect for highly boosted NIR optical performance.

# 1. INTRODUCTION

The near-infrared (NIR) region, spanning from 900 to 1700 nm, offers numerous advantages in both fundamental research and frontier applications. [1] Invisible to the naked eye, NIR light allows for the embedding of features in high-security information storage. [2] In optical communications, the use of 1550 nm has been standardized due to its low attenuation and high transparency in optical fibers, which ensures minimal energy loss during transmission. [3] In biomedicine, NIR emission wavelengths that fall within the biological imaging windows hold the potential for deep penetration owing to their low scattering and low absorption by tissues. [4] Notably, NIR-IIb (1500-1700 nm) shows reduced scattering and near-zero autofluorescence, which further enhances spatial resolution and the signal-to-noise ratio in bioimaging. [5] Additionally, NIR light is well-suited for transmitting through turbid or colored media, such as murky water, fog, and blood, and thus can be exploited for non-invasive sensing and analysis. [6]

A variety of nanoscale materials have been studied for their NIR luminescence, for example, lead sulfide/silver sulfide quantum dots, single-walled carbon nanotubes, and organic fluorophores.[1c, 7] Among these studied luminescent materials, rare-earth doped nanoparticles (RENPs) have shown great promise for their unique optical properties, such as fixed transition bandwidths, long photoluminescence lifetimes, high photostability, and low toxicity.[8] Most importantly, the abundant energy levels of the rare-earth ions allow RENPs to emit across multiple regions of the NIR spectrum, e.g., 980 nm from $Yb^{3+}$, 1060 nm and 1310 nm from $Nd^{3+}$, 1150 nm from $Ho^{3+}$, 1475 nm from $Tm^{3+}$, and 1525 nm from $Er^{3+}$. [9] In particular, $Er^{3+}$-doped RENPs, which emit in the NIR-IIb range, have been reported as promising candidates for deep tissue bioimaging and temperature sensing. [10] While a great deal of work has focused on improving the upconversion luminescence intensity [11], improving the downshifting performance, despite its various advantages and applications, is still largely unexplored. Developing bright, downshifting emissions with low excitation thresholds is still of significant importance since future applications of RENPs crucially hinge on their optical performance.

The emission in RENPs originates from the rare-earth ions doped within the host lattice.[12] In principle, a high concentration of dopant ions provides more possibilities for the absorption of excitation photons and energy transfer to the activator ions (luminescent centers) to reach a higher intensity. [13] However, excessively dense distribution of activator ions in a host can also promote concentration quenching pathways such as non-radiative energy migration to surface quenchers and non-radiative, cross-relaxation energy loss.[13-14] To date, significant efforts have been devoted to mitigating concentration quenching in RENPs. On the one hand, constructing an inert shell on the luminescent core to block the energy migration to surface quenchers is one of the most common and well-accepted approaches.[15] Nonetheless, the inert shell can only protect the activator ions from the surface quenchers while internal cross-relaxation remains a problem that hinders the high dopant concentration of activator ions. On the other hand, the integrity and heterogeneity of the widely employed core-shell structure are much clearer than before. Compelling evidence has been put forth that cation intermixing between the core and shell layer occurs such that the interfacial region is not characterized by a simple sharp separation.[16] For instance, due to the activator ions readily diffusing from the core into the shielding shell, a rather thick shell outer layer is required to properly protect the luminescent ions in the core.[17] While a great deal of research has been focused on how to suppress cation intermixing to the shielding layer,[16d, 18] utilization of the diffused elements at the interface to reach a high dopant concentration of activator ions has not yet been explored.

Here, we develop a novel approach to realize high $Er^{3+}$ doping through dopant ion diffusion at the extended interface for efficient downshifting luminescence by exploiting an inert core-active shell-inert shell architecture. Specifically, we rationally design a new structure, namely $LiLuF_4$@$LiLuF_4$:$Ce^{3+}$, $Yb^{3+}$, $Er^{3+}$@$LiLuF_4$ RENPs (**Figure 1**). To avoid host structure-induced lattice mismatch between the core and shell layers, the same composition is preferred throughout the architecture. Moreover, a homogeneous core-shell interface makes it easier for cation intermixing. A lutetium host lattice is employed compared with the more common yttrium ion host lattice given that it is better matched in size and mass with the heavy rare earth dopants.[19] Utilizing an homogeneous inert core and inert shell to construct a double-layer interface for dopant ions diffusion, the active shell layer is extended allowing for more active ions to be doped without severe concentration quenching. The size of the inert core as well as the crystal growth dynamics of the core-shell architecture have been investigated to maximize ion diffusion at the interface, which helps to better control the interface structure for enhanced downshifting luminescence intensity. By optimizing the structure, the downshifting intensity shows a dramatic enhancement compared with the conventional core-shell structure. The innovative RENPs have been successfully deployed to transmit NIR signals through obstacles including plastic and paper boxes, showcasing their exceptional capability in overcoming barriers for signal detection.

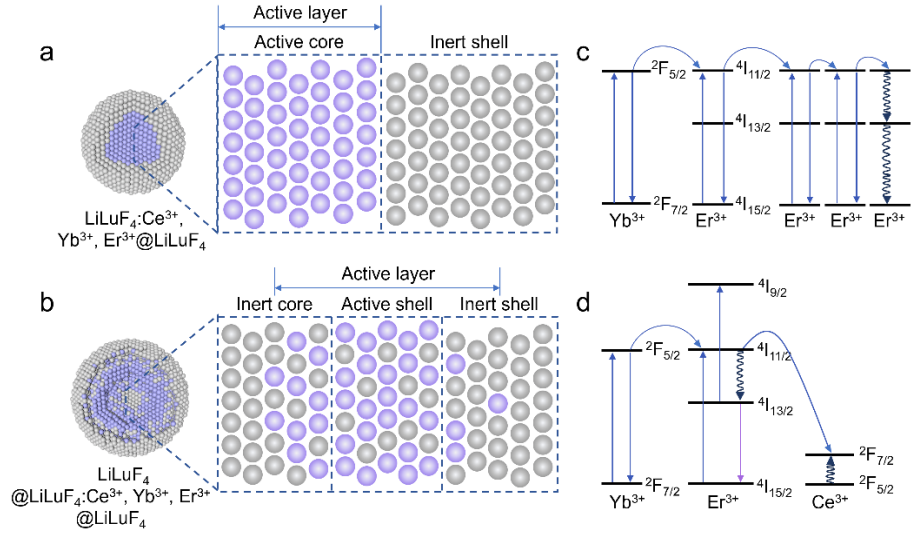

**Figure 1. Schematic of inert core-active shell-inert shell architecture with extended active layer and conventional core-shell structure.** (a) Conventional LiLuF$_4$:Ce$^{3+}$, Yb$^{3+}$, Er$^{3+}$@LiLuF$_4$ architecture with limited active layer. (b) LiLuF$_4$@LiLuF$_4$:Ce$^{3+}$, Yb$^{3+}$, Er$^{3+}$@LiLuF$_4$ inert core-active shell-inert shell structure. The extended active layer is in the middle (purple color) of the entire architecture. (c) Concentration quenching at highly Er$^{3+}$-doped conventional core-shell structure. (d) Normal energy transfer process in highly Er$^{3+}$-doped RENPs with Ce$^{3+}$ co-dopant to enhance downshifting emission intensity.

## 2. RESULTS AND DISCUSSION

A strategy of doping Ce$^{3+}$ to the matrix was used to obtain enhanced NIR downshifting emission.[1b, 10a, 20] Initially, different concentrations of Ce$^{3+}$ were doped in the LiLuF$_4$@LiLuF$_4$: $x$ mol% Ce$^{3+}$, Yb$^{3+}$, $y$ mol% Er$^{3+}$@LiLuF$_4$ ($x$ = 0, 1, 2 and $y$ = 2, 3, 4, 5, 6, 8) architecture where the concentrations of Er$^{3+}$ and Yb$^{3+}$ were fixed at 2 and 18 mol%, respectively. The morphology and size distribution were characterized by transmission electron microscopy (TEM), as shown in **Figure S1**. The inert-core (IC) with a size of $d$ nm, active-shell (AS) with dopant concentrations of Ce$^{3+}$ ($x$) or Er$^{3+}$ ($y$), and inert-shell (IS) are denoted by IC$_{d\text{nm}}$, IC$_{d\text{nm}}$-AS$_{x\text{Ce}/y\text{Er}}$, and IC$_{d\text{nm}}$-AS$_{x\text{Ce}/y\text{Er}}$-IS, respectively. To avoid size-effect induced variations on the optical performance, the same inert core with a size of about 5 nm was used and the LiLuF$_4$:$x$ mol% Ce$^{3+}$, Yb$^{3+}$, Er$^{3+}$ shell thicknesses were kept constant at about 5-8 nm. An inert LiLuF$_4$ outer shell of about 5 nm was grown as a protective layer to prevent surface quenching. All the core-shell or core-shell-shell architectures displayed a bi-pyramidal shape, indicative of the tetragonal crystal structure, which was confirmed by X-ray diffraction (XRD), as shown in **Figure S2**.

In RENPs doped with only $Yb^{3+}$ and $Er^{3+}$, the $Yb^{3+}$ ion acts as a sensitizer to harvest the excitation photons and then effectively transfer the energy to adjacent $Er^{3+}$ ions (the activator) following 980 nm excitation, Thus, the $^4I_{11/2}$ excited state of $Er^{3+}$ is subsequently populated via ground state ($^4I_{15/2}$) absorption followed by nonradiative decay to the lower lying $^4I_{13/2}$ state that ultimately leads to the 1550 nm downshifting emission ascribed to the $^4I_{13/2} \rightarrow \,^4I_{15/2}$ transition. When $Ce^{3+}$ is added to the matrix, an additional pathway accelerates the accumulation of $Er^{3+}$ ions at the $^4I_{13/2}$ level. According to the classic Dieke energy level diagram, the energy gap between the $^2F_{5/2}$ and $^2F_{7/2}$ levels of $Ce^{3+}$ is about 2300 cm$^{-1}$ (*ca.* 0.29 eV), which shows a small mismatch with the energy difference between the $^4I_{11/2}$ and $^4I_{13/2}$ states of $Er^{3+}$ (around 3700 cm$^{-1}$, *ca.* 0.46 eV).[1b, 20] Hence, when $Ce^{3+}$ is present, following the 980 nm excitation, ions in the $^4I_{11/2}$ state will easily decay to the $^4I_{13/2}$ state due to the cross-relaxation between $Er^{3+}$ and $Ce^{3+}$ ($^4I_{11/2} + \,^2F_{5/2} \rightarrow \,^4I_{13/2} + \,^2F_{7/2}$) although phonon-assisted process is necessary for the energy transfer (**Figure 1d**). [21] The influence of $Ce^{3+}$ dopant on the downshifting emission of $^4I_{11/2} \rightarrow \,^4I_{13/2}$ transition from $Er^{3+}$ under excitation of 980 nm laser is demonstrated in **Figure S3**. The multiple emission peaks within the 1550 nm manifold arise from the Stark sublevels of the $^4I_{13/2}$ and $^4I_{15/2}$ states of $Er^{3+}$ (**Figure S4**), which result from the splitting of those states due to the high crystal field strength from the $LiLuF_4$ host lattice. [10a, 22] As shown in **Figure S3**, 1 mol% $Ce^{3+}$ doped RENPs have a contrasting effect on downshifting emission enhancement independent of the architecture ($IC_{5nm}$-$AS_{xCe}$ or $IC_{5nm}$-$S_{xCe}$-S). However, a higher $Ce^{3+}$ dopant inhibits the NIR downshifting emission, indicating a limitation to the desensitization of $^4I_{13/2}$ of $Er^{3+}$. 1 mol% $Ce^{3+}$ is thus chosen as the optimum concentration for subsequent experiments.

Although the cation intermixing phenomenon at the core-shell interface has been reported from the structural perspective, there is no further demonstration of using the extended core-shell interface for higher dopant concentration and thus increased downshifting emission. [15b, 16a, 16c, 16d] To verify the dilution effect of dopant ions at the core-shell interface, the concentration of $Er^{3+}$ was tuned from 2 mol% to 8 mol%. If the diffusion phenomenon does occur at the interface, a higher dopant concentration should be tolerable without severe concentration quenching. As a proof of concept, a series of $LiLuF_4$@$LiLuF_4$:1 mol% $Ce^{3+}$, 18 mol% $Yb^{3+}$, *y* mol% $Er^{3+}$@$LiLuF_4$ were then developed. The prepared inert core-active shell $LiLuF_4$@$LiLuF_4$:1 mol% $Ce^{3+}$, $Yb^{3+}$, *y* mol% $Er^{3+}$ ($IC_{5nm}$-$AS_{yEr}$), and inert core-active shell-inert shell $LiLuF_4$@$LiLuF_4$:1 mol% $Ce^{3+}$, $Yb^{3+}$, *y* mol% $Er^{3+}$ ($IC_{5nm}$-$AS_{yEr}$-IS) RENPs are shown in **Figure 2a-b**. The size of

IC$_{5nm}$-AS$_{yEr}$ and IC$_{5nm}$-AS$_{yEr}$-IS were measured to be around 17 ± 3 nm and 27 ± 3 nm, respectively (**Figure S5**). All the RENPs were monodispersed and had a uniform size distribution, confirming the well-controlled preparation of high-quality RENPs. The radius of the core was close to the thickness of the shell, which ensured that the diameter of the optically active domain was identical when tuning the Er$^{3+}$ dopant concentration. A typical high-angle annular dark-field image in scanning transmission electron microscopy (HAADF-STEM) of IC$_{5nm}$-AS$_{yEr}$-IS is shown in **Figure 2c** with a size of 30 nm (**Figure 2d**). The high-resolution TEM (HRTEM) image displays lattice fringes with an interplanar spacing of 0.46 nm (**Figure 2e**), which matches the (101) lattice planes of tetragonal phase LiLuF$_4$ (JCPDS: 027-1251). The XRD diffractograms (**Figure S6-7**) further demonstrate the excellent crystallinity of the synthesized RENPs as well as the tetragonal phase.

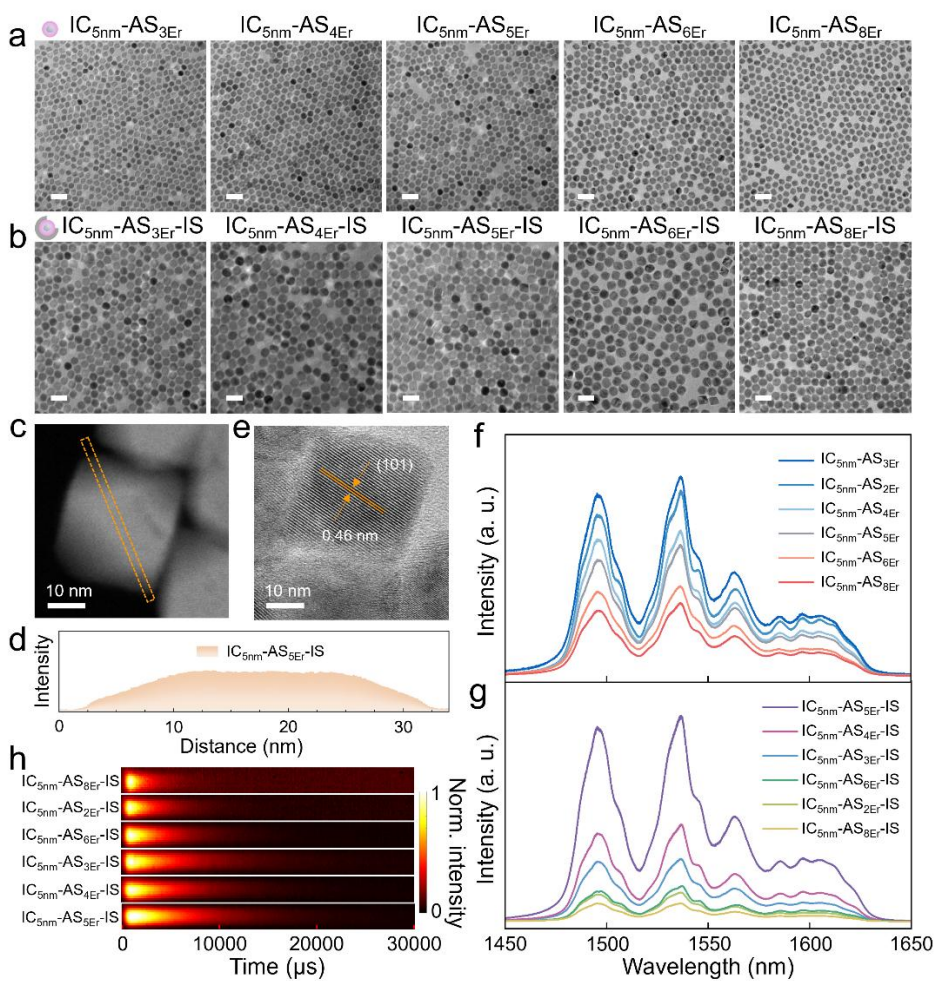

**Figure 2. Dilution effect at the core-shell interface for higher Er$^{3+}$ dopant concentrations.** TEM images of prepared RENPs with an inert core of 5 nm and varied Er$^{3+}$ dopant concentrations with the structure of (a) inert core-

active shell (IC$_{5nm}$-AS$_{yEr}$) and (b) inert core-active shell-inert shell (IC$_{5nm}$-AS$_{yEr}$-IS). (c) HAADF STEM image of a single IC$_{5nm}$-AS$_{5Er}$-IS RENP. (d) Intensity distribution along the orange bar in (c). (e) HRTEM of IC$_{5nm}$-AS$_{5Er}$-IS. (f) Downshifting spectra of IC$_{5nm}$-AS$_{yEr}$ and (g) IC$_{5nm}$-AS$_{yEr}$-IS under 980 nm laser excitation. (h) Downshifting emission intensity decay of IC$_{5nm}$-AS$_{yEr}$-IS. Scale bar in (a-b): 50 nm and $y$ = 2, 3, 4, 5, 6, 8.

**Figure 2f** shows the downshifting spectrum of the IC$_{5nm}$-AS$_{yEr}$ RENPs doped with different concentrations of Er$^{3+}$ under 980 nm laser excitation. RENPs with 3 mol% Er$^{3+}$ dopant (IC$_{5nm}$-AS$_{3Er}$) perform best among the different dopant concentrations. However, further increasing the Er$^{3+}$ dopant concentration (higher than 3 mol%) in IC$_{5nm}$-AS$_{yEr}$ resulted in decreased emission intensities. It should be noted that there is no outermost shell protection here and the emission of IC$_{5nm}$-AS$_{yEr}$ RENPs is hindered by a strong surface quenching effect. This is especially true at higher Er$^{3+}$ concentrations where more Er$^{3+}$ ions in the shell layer are closer to the surface and ultimately exposed to the environment. Therefore, the comparison of emission intensity in the IC$_{5nm}$-AS$_{yEr}$ RENPs cannot totally prove the dilution phenomenon at the interface. An outermost inert shell layer was then epitaxially grown on these RENPs. The downshifting emission spectra of IC$_{5nm}$-AS$_{yEr}$-IS are shown in **Figure 2g**. The downshifting luminescence intensity shows an initial increase from 2 mol% to 5 mol% Er$^{3+}$ dopant and then a sharp drop at 6 mol% and 8 mol%. It is noteworthy that the intensity of IC$_{5nm}$-AS$_{yEr}$-IS at $y$ = 3, 4, 5, 6 mol% are all higher than the conventional 2 mol% dopant concentration. The tolerable high dopant concentration of Er$^{3+}$ for intense downshifting emission shows direct optical evidence of the cation intermixing and thus dilution effect at the interface. However, concentration quenching still plays a dominant role when the Er$^{3+}$ concentration is too high (8 mol%). The integrated intensity of IC$_{5nm}$-AS$_{yEr}$ and IC$_{5nm}$-AS$_{yEr}$-IS at around 1550 nm is demonstrated in **Figure S8**. The enhanced downshifting intensity of IC$_{5nm}$-AS$_{yEr}$-IS compared with IC$_{5nm}$-AS$_{yEr}$ verified the successful depletion of the surface quenching. Interestingly, with the same outermost protection thickness, a significant enhancement (40 times) was realized by doping 5 mol% Er$^{3+}$ in the active shell layer while the conventional 2 mol% Er$^{3+}$ only has 4 times increment. Considering the same inert core, same active layer thickness as well as the same outermost layer thickness, the dramatic enhancement at 5 mol% Er$^{3+}$ further indicates the crucial role of the interface in high dopant concentration RENPs. Owing to the double-layer core-shell interface, Er$^{3+}$ ions dilute to both the inert core and the inert shell, so concentration quenching is weakened. The downshifting luminescence lifetimes of IC$_{5nm}$-AS$_{yEr}$-IS were characterized using an NIR luminescence lifetime imaging microscope developed in-house. [2] The time-resolved luminescence process of IC$_{5nm}$-AS$_{yEr}$-IS RENPs can be directly captured within a 30 ms exposure window in a single shot. As shown in **Figure 2h**, the intensity decay difference among all RENPs is clearly observed in the image. 5 mol% Er$^{3+}$ doped IC$_{5nm}$-AS$_{5Er}$-IS presents the longest lifetime followed by 4 mol%, 3 mol%, 6 mol%, 2 mol% and 8 mol% being the shortest one, which is consistent with the downshifting spectrum result. The normalized luminescence intensity decay curves are plotted in **Figure S9** with fitted lifetimes from 1.65 ms to 6.22 ms. The long lifetime of the $^4I_{13/2} \rightarrow \ ^4I_{15/2}$ transition indicates that the ions remain in the excited state ($^4I_{13/2}$) for a long time before decaying to the $^4I_{15/2}$ ground state, which also proves the depressed concentration quenching in 5 mol% Er$^{3+}$ highly doped IC$_{5nm}$-AS$_{5Er}$-IS.

To further reveal the effect of the interface on the optical properties, the upconversion emission spectra of IC$_{5nm}$-AS$_{yEr}$-IS were also studied in **Figure S10**. All the samples presented two characteristic visible emission bands centered at 525/545 and 660 nm, which are attributed to the radiative transitions of the $^2H_{11/2}/^4S_{3/2}$, and $^4F_{9/2}$ excited states to the $^4I_{15/2}$ ground state, respectively. The results revealed that the 5% Er$^{3+}$ doped IC$_{5nm}$-AS$_{5Er}$-IS sample, still possessed the highest upconversion emission intensity among the samples studied. Nonetheless, the difference in the upconversion emission intensity among the varied Er$^{3+}$ dopant concentrations was not as obvious as in the downshifting emission. It is noted that upconversion and downshifting emission intensities are highly related to the population of the $^4I_{11/2}$ excited state where the two processes (i.e., upconversion or downshifting) are actually in competition with each other. In our case, the addition of Ce$^{3+}$ promoted phonon-assisted energy transfer from $^4I_{11/2}$ of Er$^{3+}$ to Ce$^{3+}$ resulting in the inevitable suppression of the upconversion emission.

In addition to the dopant concentration, it is also meaningful to explore the influence of the existence of the inert core and its size on the interface dilution effect. It has been reported that the size of the starting core can change the diffusion length to affect the cation intermix . [16a, 23] Here, we tuned the inert core from 5 nm to 10 nm. A series of RENPs with an inert core of 10 nm (IC$_{10nm}$), active-shell layers doped with 2 mol% Er$^{3+}$ (AS$_{2Er}$), and 5 mol% Er$^{3+}$ (AS$_{5Er}$) are shown in **Figure 3b-c**. The size (**Figure S11**) of the active shell, and outermost shell were kept similar at around 5 nm. Besides, the conventional active core and core-shell structures are shown in **Figure 3d** for comparison. To ensure the same diameter of the optically active domain in the final structure, the thickness of the active shell in IC$_{5nm}$-AS$_{yEr}$-IS and IC$_{10nm}$-AS$_{yEr}$-IS was set to be as close as the radius of the conventional core. The size difference between the two adjacent architectures only came from the inert core and was thus expected to be 5 nm (as confirmed in **Figure 3e**). Furthermore, the clear lattice fringes observed over the entire architecture in the HRTEM images of the $AC_{2Er}$-IS, $IC_{5nm}$-$AS_{2Er}$-IS, and $IC_{10nm}$-$AS_{2Er}$-IS samples (**Figure S12**) demonstrate their excellent crystallinity and the tetragonal phase (**Figure S13**).

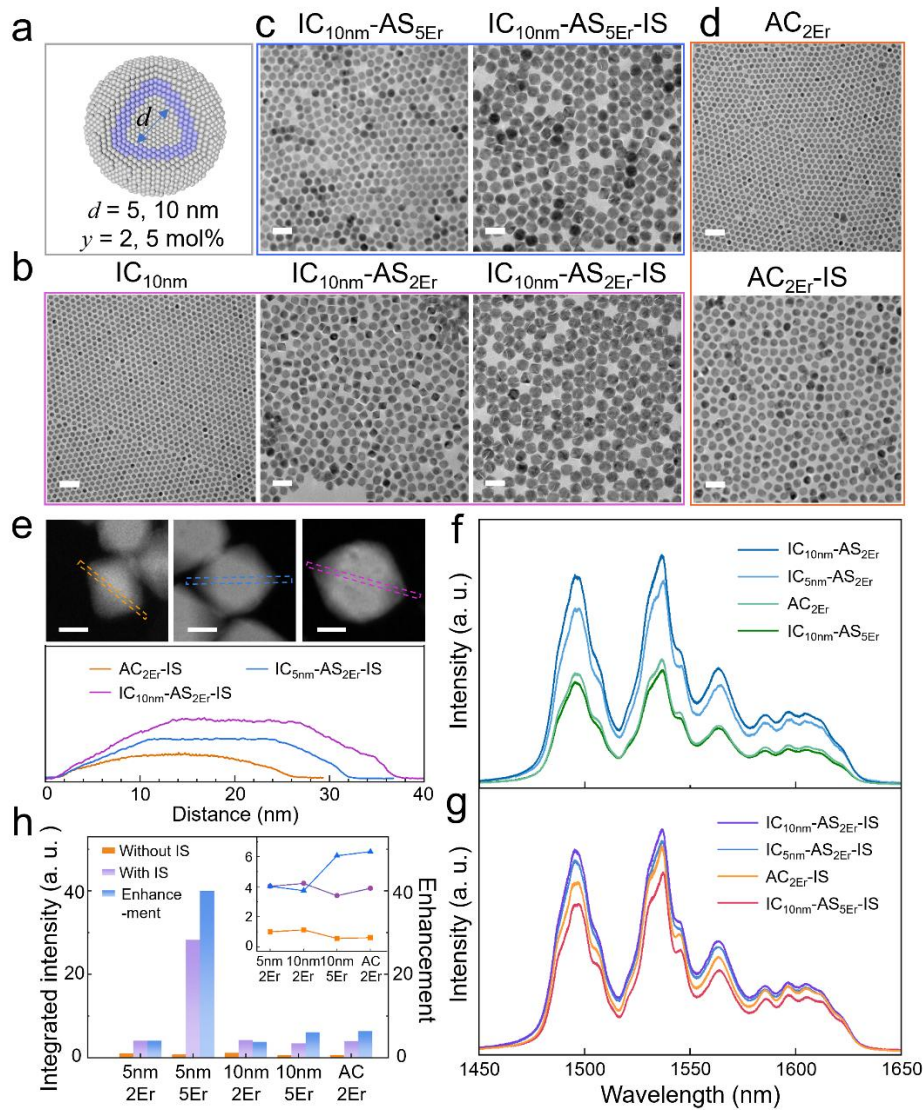

**Figure 3. Influence of the inert core on the downshifting emission intensity.** (a) Schematic of the inert core-active shell-inert shell architecture. (b-d) TEM images of explored RENP structures: (b) inert core ($IC_{10nm}$), inert core-active shell ($IC_{10nm}$-$AS_{2Er}$), and inert core-active shell-inert shell ($IC_{10nm}$-$AS_{2Er}$-IS) with an inert core of 10 nm and 2 mol% $Er^{3+}$ dopant concentration; (c) inert core-active shell ($IC_{10nm}$-$AS_{5Er}$) and inert core-active shell-inert shell ($IC_{10nm}$-$AS_{5Er}$-IS) with an inert core of 10 nm and 5 mol% $Er^{3+}$ dopant concentration; (d) active core ($AC_{2Er}$) and active core-inert shell ($AC_{2Er}$-IS) with 2 mol% $Er^{3+}$ dopant concentration. (e) HAADF-STEM images of $AC_{2Er}$-IS,

IC$_{5nm}$-AS$_{2Er}$-IS, IC$_{10nm}$-AS$_{2Er}$-IS, and the corresponding averaged intensity distribution along the selected area. Downshifting emission spectra of (f) AC$_{2Er}$ and IC$_{dnm}$-AS$_{yEr}$ with $d$ = 5, 10 nm and $y$ = 2 mol% and 5 mol%, (g) similar to (f) but the structure has the outermost inert shell coating. (h) Integrated NIR downshifting intensity in (f-g) and the enhancement factor of the outermost shell protection. The inset is a magnified plot of (h) without the 5nm/5Er set. Scale bar: 50 nm in (a-b), 50 nm in (e).

The downshifting spectra of RENPs without the outermost shell layer are shown in **Figure 3f**. For IC$_{dnm}$-AS$_{yEr}$ with the same Er$^{3+}$ concentration but different size of the inert core, IC$_{10nm}$-AS$_{2Er}$ shows higher emission intensity than IC$_{5nm}$-AS$_{2Er}$ and AC$_{2Er}$ due to its larger size, smaller surface-to-volume ratio and thus relatively fewer surface quenching centers. For IC$_{dnm}$-AS$_{yEr}$ with different Er$^{3+}$ concentrations but the same inert core, the downshifting intensity of IC$_{10nm}$-AS$_{5Er}$ is lower than IC$_{10nm}$-AS$_{2Er}$, indicating the strong concentration quenching in IC$_{10nm}$-AS$_{5Er}$. After outermost-shell protection, the intensity order in **Figure 3g** is the same as it is in **Figure 3f** without the outermost-shell. The integrated intensities of the RENPs are normalized according to IC$_{5nm}$-AS$_{2Er}$ (**Figure 3h**). Surprisingly, despite the significantly high intensity of IC$_{5nm}$-AS$_{5Er}$-IS, IC$_{10nm}$-AS$_{5Er}$-IS shows even lower intensity than the conventional core-shell structure, which means the contribution of double layer interface for enhancing the NIR intensity in IC$_{10nm}$-AS$_{5Er}$-IS is limited (**Figure S14-15**) and also the dilution effect at the interface is not obvious.

To explore the possible influence of the crystal structure of the starting cores on the NIR downshifting emission intensity, we performed energy-dispersive X-ray spectroscopy (EDS) line-scanning for AC$_{2Er}$-IS, IC$_{5nm}$-AS$_{2Er}$-IS and IC$_{10nm}$-AS$_{2Er}$-IS (**Figure S16**). Under identical measurement conditions, the IC$_{5nm}$-AS$_{2Er}$-IS sample was damaged after line scanning while AC$_{2Er}$-IS and IC$_{10nm}$-AS$_{2Er}$-IS kept their original morphology. The elemental distribution of Lu and F both present a decreasing trend in the center of IC$_{5nm}$-AS$_{2Er}$-IS crystal indicating that the damage mainly occurs at the center of the crystal. Following this, the same characterization condition was also conducted on IC$_{5nm}$. Compared with the HAADF image before line scanning, the IC$_{5nm}$ crystal shows obvious damage after line scanning (**Figure S18a-b**), which can also be observed in the HRTEM image (**Figure S18d**) where two circled RENPs show the variation. The fragile structure of IC$_{5nm}$ suggests its instability under high-energy electron beam irradiation in HAADF. The crystallinity of IC$_{5nm}$ was then investigated by XRD (**Figure S17**). First, the diffraction peaks of the core match well with the tetragonal phase LiLuF$_4$ regardless of its size.

However, the peak patterns of IC$_{5nm}$ are not sharp and narrow compared with the AC$_{10nm}$ and IC$_{10nm}$, implying the poor crystallinity of IC$_{5nm}$. The instability and low crystallinity of such ultra-small RENPs have also been observed in previous research. [24] Moreover, in a typical synthesis procedure, stabilization of the 5 nm-sized core is necessary to get a chemically stable structure. [24] However, in our specific case, the stabilization step of the 5 nm inert core was bypassed to obtain the required small size.

From the downshifting emission intensity and structure properties, we determined that a smaller un-stabilized inert core, IC$_{5nm}$, was easier for cation intermixing at the interface, which was attributed to its instability and poor crystallinity. Previous reports have shown that 8 nm starting cores (in NaEuF$_4$@NaGdF$_4$) demonstrated more significant cation intermixing than starting cores of 40 nm (in NaYF$_4$:Ce@NaYF$_4$:Tb). [15b, 25] Meanwhile, Hudry *et al.* discussed the interface structure in detail, concluding that smaller starting cores had a higher probability of losing the chemical integrity of the starting core domain. [16c, 26] On the contrary, while the larger inert core (IC$_{10nm}$) presented more stable crystallinity, it was more difficult for the diffusion process to occur and thus a well-defined and abrupt interface was prone to be created. The highly doped Er$^{3+}$ ions in IC$_{10nm}$-AS$_{5Er}$-IS were confined in the middle layer, so the internal cross-relaxation was more severe, which was not ideal for the downshifting emission intensity.

For a better understanding of the diffusion process on the downshifting emission intensity, IC$_{5nm}$-AS$_{5Er}$ RENPs were extracted at different reaction times (i.e., 10, 25, 40, and 60 min) after the hot injection of the shell precursors (**Figure 4a**). The downshifting emission spectra and integrated intensity (**Figure 4e, g**) show that IC$_{5nm}$-AS$_{5Er}$ at 25 min performs better. The same phenomenon was also observed in the RENPs with a lower Er$^{3+}$ doping concentration in the active shell, IC$_{5nm}$-AS$_{2Er}$ (**Figure S19**). The sizes of IC$_{5nm}$-AS$_{5Er}$ at different reaction times (**Figure 4c**) were similar, indicating that the formation of the LiLuF$_4$ shell was quite fast. However, a subtle decreasing trend can also be seen from 10 min to 25 min, indicating the dissolution of the architecture. At the same time, the narrowed size distribution at 40 min and 60 min implies the re-growth to more uniform morphology during the core-shell synthesis. Heating of RENPs is known to induce the Ostwald ripening process in which small RENPs are dissolved (from 10 min to 25 min) into monomers until the concentration of the monomers in solution becomes so high that the particles stop dissolving and start to grow (from 25 min to 60 min). The consumed monomers in the solution will be further replenished to keep the thermodynamic balance.[27] In consequence, the dissolution of IC$_{5nm}$ at 25 min was expected to be high and gradually decrease. Er$^{3+}$ ions were redistributed in the architecture with a lower concentration in the active shell and thus the concentration quenching is suppressed. Similar results have also been observed in the literature: e.g., a low concentration of Eu$^{3+}$ ions is found in the shell layer of the NaEuF$_4$@NaGdF$_4$ architecture, Yb$^{3+}$ ions are spatially diluted within the whole volume of the NaYF$_4$:Yb$^{3+}$, Er$^{3+}$@NaYF$_4$ RENP according to the EDS maps. [15b, 28] In the case of IC$_{5nm}$-AS$_{2Er}$-IS, the inert shell should first dissolve into the active shell to dilute the concentration of Er$^{3+}$ ions, followed by growth with high crystallinity to block the surface quenchers. In **Figure 4d**, 40 min reaction time with a narrow size distribution was the expected reaction time, and the NIR emission intensity (**Figure 4f**, **g**) further proved the expectation.

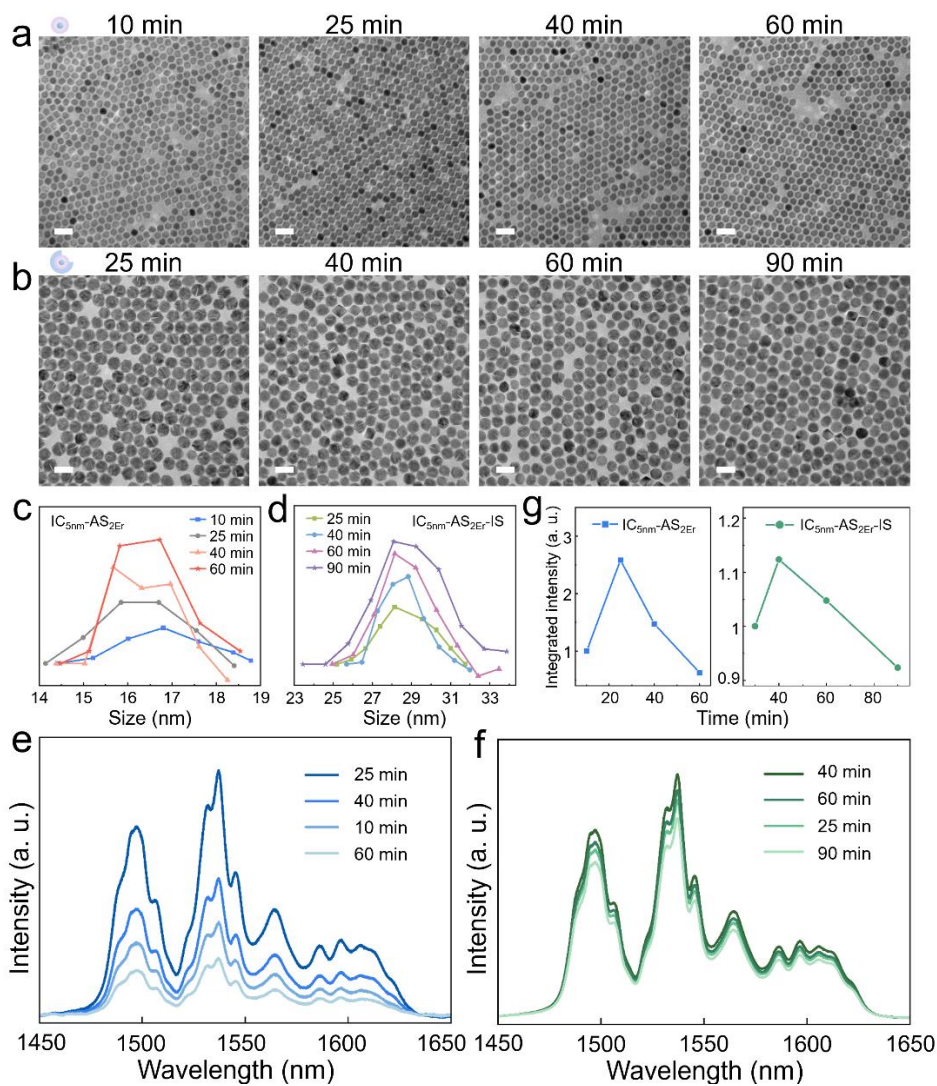

**Figure 4. Morphology and downshifting emission intensity of IC$_{5nm}$-AS$_{5Er}$ and IC$_{5nm}$-AS$_{5Er}$-IS extracted at different reaction times during the shell growth.** (a-b) TEM images of (a) IC$_{5nm}$-AS$_{5Er}$ reacted at 10, 25, 40, 60 mins and (b) IC$_{5nm}$-AS$_{5Er}$-IS extracted at 25, 40, 60, 90 mins. (c-d) Size distribution of the RENPs in the images in (a-b). (e) Downshifting emission spectra of the samples in (a). (f) As (e) but for the samples in (b). (g) Normalized integrated intensity of samples in (a-b). Scale bar in (a-b): 50 nm.

By finely tuning the parameters including the size of the inert core, the dopant concentration of Er$^{3+}$, and the reaction time of shell growth, the downshifting luminescence intensity of these newly designed IC$_{5nm}$-AS$_{5Er}$-IS RENPs is 12 times higher than the conventional core-shell structure AS$_{2Er}$-IS (**Figure S20**). It is noteworthy that the main contribution to this remarkable 12-times enhancement is the 5 nm inert core-active shell-inert shell architecture which allows for a high (5 mol%) Er$^{3+}$ high dopant concentration. To give a direct impression of enhancement in the NIR intensity, the NIR images of newly developed IC$_{5nm}$-AS$_{5Er}$-IS RENPs and conventional AC$_{2Er}$-IS RENPs covered with an "INRS" transparency are shown in **Figure 5a-b**. The different optical performance in the NIR intensity as well as the lifetime of IC$_{5nm}$-AS$_{5Er}$-IS and AC$_{2Er}$-IS is further demonstrated by lifetime imaging microscopy [2]. We first mixed the two samples as follows: hexane solution containing AC$_{2Er}$-IS RENPs was deposited on the glass slide in a relatively large area. Using the same protocol, IC$_{5nm}$-AS$_{5Er}$-IS hexane solution was then dropped at the center of the same area. In **Figure 5c**, brighter NIR intensity along with the longer lifetime in the middle of the mixture comes from the emission of IC$_{5nm}$-AS$_{5Er}$-IS while the surrounding part is mainly the distribution of AC$_{2Er}$-IS RENPs. To showcase the significance of the enhanced NIR downshifted intensity in signal detection, we applied the optimized IC$_{5nm}$-AS$_{5Er}$-IS RENPs to investigate their capabilities to transmit their NIR luminescence through (diverse) obstacles for signal detection. As shown in Fiugre 5d**,** the visible image presents no useful information underneath the obstacle, cellulose paper, while the NIR image (**Fiugre 5e**) depicts the "A" letter which is the transparency covered on IC$_{5nm}$-AS$_{5Er}$-IS RENPs. In contrast, nearly no hint of the shape of the letterunderneath the obstacles can be obtained from the NIR image (**Fiugre 5f**) of conventional structure, AC$_{2Er}$-IS RENPs. The low intensity of AC$_{2Er}$-IS induced low signal-to-noise ratio so that the useful information was submerged into the background (**Fiugre 5g**). The capilibltiy of the high NIR intensity of IC$_{5nm}$-AS$_{5Er}$-IS RENPs was further demonstrated using diverse opaque obstacles (**Figure S21**): polyethylene terephthalate; polypropylene with acrylic adhesive; cellulose paper. Without covering any obstacles, a clear image with sharp boundaries of the covered "C", "A", and "N" letters could be observed (**Figure S21a**). The uneven intensity distribution is from the height difference of the powder sample so that only part of the area was focused. By covering diverse opaque obstacles: polyethylene terephthalate (**Figure S21e**); polypropylene with acrylic adhesive (**Figure S21f**); cellulose paper (**Figure S21g**), the image of "C" "A" "N" letters are blurred with decreased intensity (**Figure S21b-d**) which mainly due to the scattering and absorption of the medium. Nonetheless, the shape can still be seen to discern the pattern even when the signal is low and the covered material has different colors (**Figure S22**). The newly designed IC$_{5nm}$-AS$_{5Er}$-IS with the dramatic 12 times enhanced NIR intensity shows its irreplaceable role in signal detection.

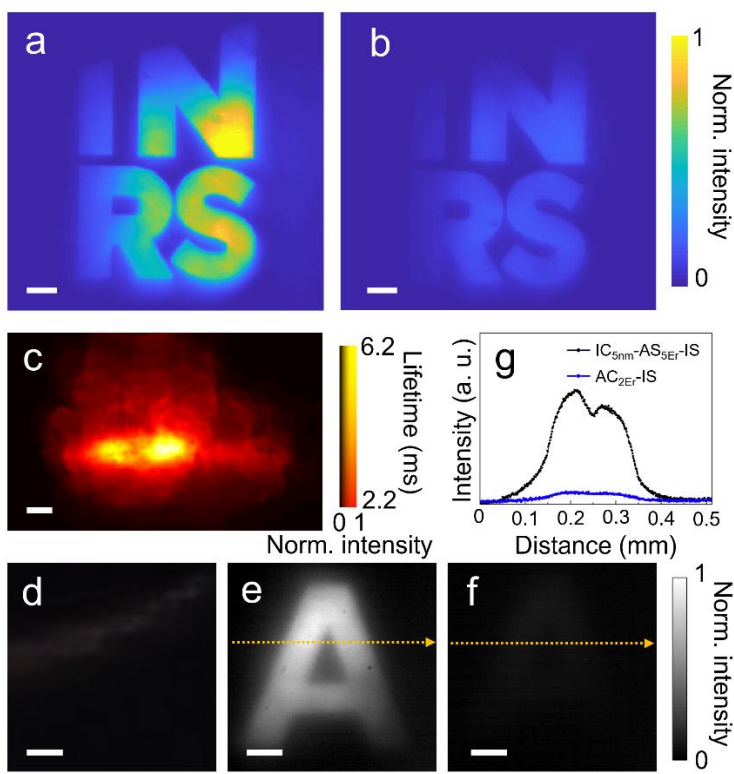

**Figure 5. Enhanced NIR intensity of IC$_{5nm}$-AS$_{5Er}$-IS and its application in signal detection through obstacles.** Normalized NIR intensity image of (a) IC$_{5nm}$-AS$_{5Er}$-IS and (b) AC$_{2Er}$-IS under 980 nm excitation. (c) NIR intensity and lifetime distribution of the mixed sample of IC$_{5nm}$-AS$_{5Er}$-IS and AC$_{2Er}$-IS. (d) Bright field image and NIR intensity image of the (e) IC$_{5nm}$-AS$_{5Er}$-IS and (f) AC$_{2Er}$-IS covered with cellulose paper. (g) Intensity profile of the marked place in (e and f). Scale bar: 50 μm.

## 3. CONCLUSION

In summary, we have developed a novel RENP architecture, based on an inert core-active shell-inert shell structure, specifically LiLuF$_4$@LiLuF$_4$:Ce$^{3+}$, Yb$^{3+}$, Er$^{3+}$@LiLuF$_4$ in which the Er$^{3+}$ dopant concentration can be increased to 5% yielding 12 times enhanced NIR downshifting emission intensity at around 1550 nm compared to conventional active core-inert shell architecture. The enhancement mechanism was further studied by tuning the size/crystallinity of the inert core (5 nm and 10 nm) as well as the shell growth dynamics. The ultra-small unstable inert core allowed for easier intermixing with the shell layers. Meanwhile, the double interfaces in the LiLuF$_4$@LiLuF$_4$:Ce$^{3+}$, Yb$^{3+}$, Er$^{3+}$@LiLuF$_4$ core-shell-shell architecture provided an extended layer for Er$^{3+}$ doping to overcome the concentration quenching effect that normally occurs in the conventional core-shell structure. Our method suggests a general way to suppress concentration quenching to further increase not only NIR downshifting emission intensity but also visible emissions. The optimized structure IC$_{5nm}$-AS$_{5Er}$-IS has been successfully applied in signal detection through opaque obstacles.

## 4. EXPERIMENTAL METHODS

*Preparation of RENPs precursors.* Stoichiometric amounts of RE$_2$O$_3$ (RE = Lu, Yb, Er) and Ce$_2$(CO$_3$)$_3$ were selectively mixed with 5 mL trifluoroacetic acid and 5 mL distilled water in a 50 mL three-neck round bottom flask. The mixture was kept at 80 °C under vigorous stirring until the solution became clear. The temperature was then reduced to 60 °C to evaporate the residual trifluoroacetic acid and water.

*Synthesis of 5 nm LiLuF$_4$.* Unstabilized LiLuF$_4$ core RENPs were synthesized via the hot injection thermolysis approach. Solution A0: 7 mL of oleic acid (OA), 7 mL of oleylamine (OM), and 14 mL 1-octadecene (ODE), were degassed at 110 °C for 15 min and then heated to 330 °C under an argon atmosphere. Solution B0: 2.5 mmol CF$_3$COOLi and (CF$_3$COO)$_3$RE (RE = Lu/Yb, Er) were mixed with 3 mL OA and 6 mL ODE, and was degassed at 125 °C for 30 min. 3 mL of OM was added in Solution B0 and degassed for 5 min. Solution B0 was then injected into Solution A0 with a rate of 1.5 mL min$^{-1}$. After reacting at 330 °C for 1 h, the preparation of ultra-small LiLuF$_4$ was finished.

*Synthesis of 10 nm LiLuF$_4$ and conventional core LiLuF$_4$:1 mol% Ce$^{3+}$, 18 mol% Yb$^{3+}$, 2 mol% Er$^{3+}$.* Core RENPs were formed by the stabilization of the first nuclei, ultra-small LiLuF$_4$, with an excess of OA. 1.25 mmol first nuclei were mixed with 16 mL of OA and 16 mL of ODE in a 100 mL three-neck round bottom flask. The solution was degassed at 110 °C for 30 min and backfilled with argon gas. The temperature was raised to 315 °C following which the reaction was continued for 1 h. conventional core LiLuF$_4$:1% Ce$^{3+}$, 18% Yb$^{3+}$, 2% Er$^{3+}$ was synthesized using same procedure as 10 nm LiLuF$_4$ with extra addition of Yb$^{3+}$, Er$^{3+}$, and Ce$^{3+}$ source in Solution B.

*Synthesis of LiLuF$_4$@ LiLuF$_4$:x mol% Ce$^{3+}$, 18 mol% Yb$^{3+}$, y mol% Er$^{3+}$@LiLuF$_4$ (x = 0, 1, 2; y = 2, 3, 4, 5, 6, 8).* The core-shell structure was prepared by epitaxial growth of the shell on the performed cores. 0.5 mmol of LiLuF$_4$ inert core RENPs (size of the inert core: 5/10 nm) were mixed with 10 mL of OA and 10 mL of ODE (Solution A1). 2 mmol shell precursors (CF$_3$COO)$_3$RE (RE = Lu/Yb/ Er/Ce) together with CF$_3$COOLi were mixed with 10 mL of OA and 10 mL of ODE (Solution B1). Solution A1 and B1 were both degassed at 110 °C for 30 min. After degassing, Solution A1 was heated to 315 °C. Solution B1 was injected into Solution A1 at a 1.5 mL min$^{-1}$ rate when the temperature of Solution A1 was stable. After cooling down to room temperature, the LiLuF$_4$@LiLuF$_4$:1 mol% Ce$^{3+}$, 18 mol% Yb$^{3+}$, 2 mol% Er$^{3+}$ inert core-active shell RENPs were washed with hexane/ethanol (1:3) three times and re-dispersed in hexane. LiLuF$_4$@LiLuF$_4$:1% Ce$^{3+}$, 18% Yb$^{3+}$, 2% Er$^{3+}$@LiLuF$_4$ inert core-active shell-inert shell RENPs were prepared by epitaxial growth of the LiLuF$_4$ shell onto the inert core-shell structure. Conventional core-shell structure LiLuF$_4$:1 mol% Ce$^{3+}$, 18 mol% Yb$^{3+}$, 2 mol% Er$^{3+}$@LiLuF$_4$ was synthesized following the same protocol.

**Supporting Information**

Supporting Information is available from the Wiley Online Library or from the authors.

**Acknowledgments**

This work was supported in part by Natural Sciences and Engineering Research Council of Canada (RGPIN-2024-05551, I2IPJ 592655-24), Canada Research Chair Programs (CRC-2022-00119), INRS Chair in Nano-biophotonics, Canadian Cancer Society (707056); New Frontier in

Research Fund (NFRFE-2020-00267); Fonds de Recherche du Québec–Nature et Technologies (203345–Centre d'Optique, Photonique, et Lasers); Fonds de Recherche du Québec–Santé (267406,280229).

**Conflict of Interest**

The authors declare no conflict of interest.

**Data Availability Statement**

The data that support the findings of this study are available from the corresponding authors upon reasonable request.

**Keywords**

rare-earth doped nanoparticles, efficient downshifting emission, cation intermixing, core-shell interface, signal detection